\documentclass{JINST}
\usepackage{graphicx}

\title{Modeling an array of encapsulated germanium detectors}

\author{Ritesh Kshetri$^a$\thanks{Corresponding
author.}~\\
\llap{$^a$}Saha Institute of Nuclear Physics,\\
  1/AF, Bidhannagar, Kolkata - 700064, India.\\

  E-mail: \email{ritesh.khetri@saha.ac.in}}

\abstract{A probability model has been presented for understanding the operation of an array of encapsulated germanium detectors generally known as composite detector. The addback mode of operation of a composite detector has been described considering the absorption and scattering of $\gamma$-rays. Considering up to triple detector hit events, we have obtained expressions for peak-to-total and peak-to-background ratios of the cluster detector, which consists of seven hexagonal closely packed encapsulated HPGe detectors. Results have been obtained for the miniball detectors comprising of three and four seven hexagonal closely packed encapsulated HPGe detectors. The formalism has been extended to the SPI spectrometer which is a telescope of the INTEGRAL satellite and consists of nineteen hexagonal closely packed encapsulated HPGe detectors. This spectrometer comprises of twelve detector modules surrounding the cluster detector. For comparison, we have considered a spectrometer comprising of nine detector modules surrounding the three detector configuration of miniball detector. In the present formalism, the operation of these sophisticated detectors could be described in terms of six probability amplitudes only. Using experimental data on relative efficiency and fold distribution of cluster detector as input, the fold distribution and the peak-to-total, peak-to-background ratios have been calculated for the SPI spectrometer and other composite detectors at 1332 keV. Remarkable agreement between experimental data and results from the present formalism has been observed for the SPI spectrometer.}


\begin{document}

\section{Introduction}

In gamma-ray spectroscopy, a novel way of obtaining high full energy peak (FEP)
\footnote{The full energy peak (FEP) corresponds to complete absorption of incident gamma energy. For gamma-rays with energies from 10 keV to 10 MeV, three types of interactions are important for gamma-ray detection: the photoelectric absorption, the Compton scattering and the pair production (possible when gamma-ray energy $\ge$ 1.022 MeV) \cite{ge}. FEP gets contributions from photoelectric absorption in a single step and from the following processes in multiple steps - (1) Compton scattering where the scattered gamma-ray is absorbed by photoelectric effect and (2) pair production where both the annihilating 511 keV gamma-rays are absorbed by photoelectric effect. In some cases, the high energy Compton scattered gamma-rays can produce pair production. There will also be events where the annihilating 511 keV gamma-rays are Compton scattered. All such events contribute to FEP when all the scattered photons are eventually absorbed in the medium by photoelectric effect.} 
detection efficiency without deteriorating the energy resolution and timing characteristics is the use of composite detectors \cite{ge} which are composed of standard high purity germanium (HPGe) crystals arranged in a compact way. Two simple examples are the clover and cluster detectors. The clover detector \cite{ge,clov} consists of four closely packed high purity germanium (HPGe) crystals (having tapered square structure) inside the same cryostat, while the cluster detector \cite{ge,ebe} consists of seven closely packed hexagonal encapsulated HPGe detectors inside the same cryostat. Other examples of arrays or groups of encapsulated detectors include the miniball detector array and the SPI spectrometer. The Miniball detector array \cite{ge,mini} consists of eight cryostats with three closely packed hexagonal encapsulated HPGe detectors in each plus four cryostats with four closely packed hexagonal encapsulated HPGe detectors in each. Here each HPGe detector is electrically segmented into ($6\times1$) segments. The SPI spectrometer \cite{spi,spi2} comprises of nineteen closely packed hexagonal encapsulated HPGe detectors. This spectrometer is a telescope of the INTEGRAL satellite and addresses the fine spectroscopy of celestial $\gamma$-ray sources in the energy range of 20 keV to 8 MeV. Except the clover detector which will not be considered in this text, for all other detectors mentioned above, the HPGe crystals inside encapsulated modules have a hexagonal shape with a side length of 3.2 cm and a height of 7 cm. The central bore is 6 mm in diameter and 6 cm in length. Each HPGe crystal is individually mounted inside a partially pressurized aluminum capsule.

Let us consider a general composite detector consisting of {\it K} closely packed hexagonal encapsulated HPGe detector modules. If a $\gamma$-ray is incident on the composite detector, then the interaction of the $\gamma$-ray with one of the detector modules can cause time correlated events consisting of scattered $\gamma$-rays to adjacent detectors. If we add up these time correlated events, then information from the scattered $\gamma$-rays which do not escape from the composite detector, will be added back to the full energy peak (FEP). As a result, the composite detector can be operated in two modes - (i) the single detector mode which is the time uncorrelated sum of data from {\it K} detector modules corresponding to events where the full $\gamma$-ray energy is  deposited in any one of the individual detectors, and (ii) the addback mode which is the time correlated sum of data from {\it K} detector modules. The later mode corresponds to events where the full $\gamma$-ray energy is deposited by single and multiple hits. Due to these multiple hit events, there are more full energy peak counts and less background 
\footnote{The background (including escape peaks) corresponds to incomplete absorption of incident gamma energy \cite{ge}. It corresponds to events where scattered gamma-ray(s) escape the detector after partial energy deposition. As an example, if after single or multiple Compton scattering, scattered gamma-ray with reduced energy escapes from the detector, then the event contributes to the background of the spectrum. Similarly, if one of the 511 keV gamma-ray escapes, then a peak is observed at $E_{g} - 511$ (single escape peak) and if both escape, then a peak is observed at $E_{g} - 2\times511$ (double escape peak), where $E_{g}$ is the energy of the incident gamma-ray in keV.}
(including escape peaks) counts in addback mode \cite{ge,clov}. As a result, the FEP efficiency and peak-to-total ratio are higher in addback mode compared to single detector mode. Figure 1 shows schematic diagrams of the various composite detectors consisting of {\it K} closely packed hexagonal encapsulated HPGe detectors. It can be observed that the SPI spectrometer comprises of twelve detector modules surrounding the cluster detector. Similarly, here we have considered a spectrometer comprising of nine detector modules surrounding the three detector configuration of miniball detector. Since this composite detector has twelve detectors, we have named it - {\it K}12 detector. The main purpose of considering the {\it K}12 detector is to demonstrate the power of the formalism to predict the response of composite detectors similar to the cluster detector.

In the present paper, we have tried to describe the addback mode of operation of five sophisticated composite detectors considering absorption and scattering of gamma-rays. The expressions for fold distribution and peak-to-background, peak-to-total ratios have been obtained.

\section{The model and its applications}

We assume that inside the composite detector, the HPGe hexagonal detector modules have identical shape, size and are placed symmetrically. Let $N_T$ be the total flux of monoenergetic $\gamma$-rays incident on the composite detector such that at a time a single $\gamma$-ray could interact with a detector module. Let $N$ be the portion of the total flux that interact with a detector module of the composite detector. For these $N$ $\gamma$-rays (of energy $E_{\gamma}$) incident on a detector module, let the probability of scattering away from the detector without detection be $S'_o$, the probability of scattering to adjacent detectors be $S'_i$ and the probability of full energy peak absorption be $A'$, such that $S'_o + A' + S'_i =$ 1. Note that after first interaction of $N$ $\gamma$-rays with a detector module, $NS'_i$ are scattered to adjacent detectors. 

For each detector module inside the composite detector, the probability of scattering in and out are the same for adjacent detectors, and the probability of a $\gamma$-ray to traverse from one detector to another after scattering is also the same. Each detector can either absorb (with full energy peak absorption probability $A$) or scatter a $\gamma$-ray away from its volume (with probability $S_o$ for leaving detector without detection and probability $S_i$ for scattering to adjacent detectors), such that $A + S_o + S_i =$ 1. Due to the identical shape, size of the detectors and the symmetric configuration, $A$, $S_i$, $S_o$ are same for all detectors. However, these probabilities are different from $A', S'_i$ and $S'_o$ which are probabilities for first interaction. Later during multiple scattering, if scattered $\gamma$-ray again enters the detector of first interaction, then its absorption and scattering will be described by $A, S_i$ and $S_o$. Note that $A', S'_i, S'_o, A, S_i, S_o$ are probabilities integrated over energies and angles of scattered gamma-rays.

Let us consider an event where a gamma-ray interacts with a composite detector and is absorbed, thereby contributing to the FEP. The energy of the gamma-ray could be deposited completely in a single detector module (corresponding to single fold
\footnote{For a composite detector operated in addback mode, the energies deposited in several detectors because of Compton scattering and or pair production can be added up and the full energy of a $\gamma$-ray can be determined in many cases. The number of individual detector modules that participate to one Compton scattering and or pair production event is called "fold".}
) or there is partial deposition of gamma energy in $m$ modules (corresponding to multiple fold, $2 \le m \le 4$). Thus, a FEP event can generate $m$ background counts in spectrum for single detector mode of operation. However, in the spectrum corresponding to addback mode, the $m$ background counts are reconstructed to a single FEP count.  So, corresponding to each FEP event, there is a single FEP count in addback mode. So, for the addback mode, the number of events contributing to FEP is equal to counts under the FEP in the spectrum. In the present formalism, for simplicity, we will consider the case of addback mode.

Experimental fold distribution of the cluster detector have shown that triple fold events contribute $\approx$ 3\% to full energy peak events for gamma-rays of energy 1.3 MeV \cite{wil}. It is observed that the contribution from four and higher fold events is negligible. As a result, here the addback mode will consist of up to three fold events and we will neglect the contribution of four and higher fold events. Let the full energy peak, background and total counts in addback mode be denoted by $P$, $B$ and $T$, respectively. For each of the composite detectors schematically shown in figure 1, we will now consider the expression of total counts for finding the number of escaped and absorbed events. We will first consider the cases for {\it K =} 7, 19 and thereafter {\it K =} 3, 4 and 12. For each case, we will first consider the total events to find the peak and background counts (note that for addback mode, number of events = number of counts).

\begin{figure}[htp]
\centering
\includegraphics[totalheight=0.49\textheight,viewport=43 260 780 795,clip]{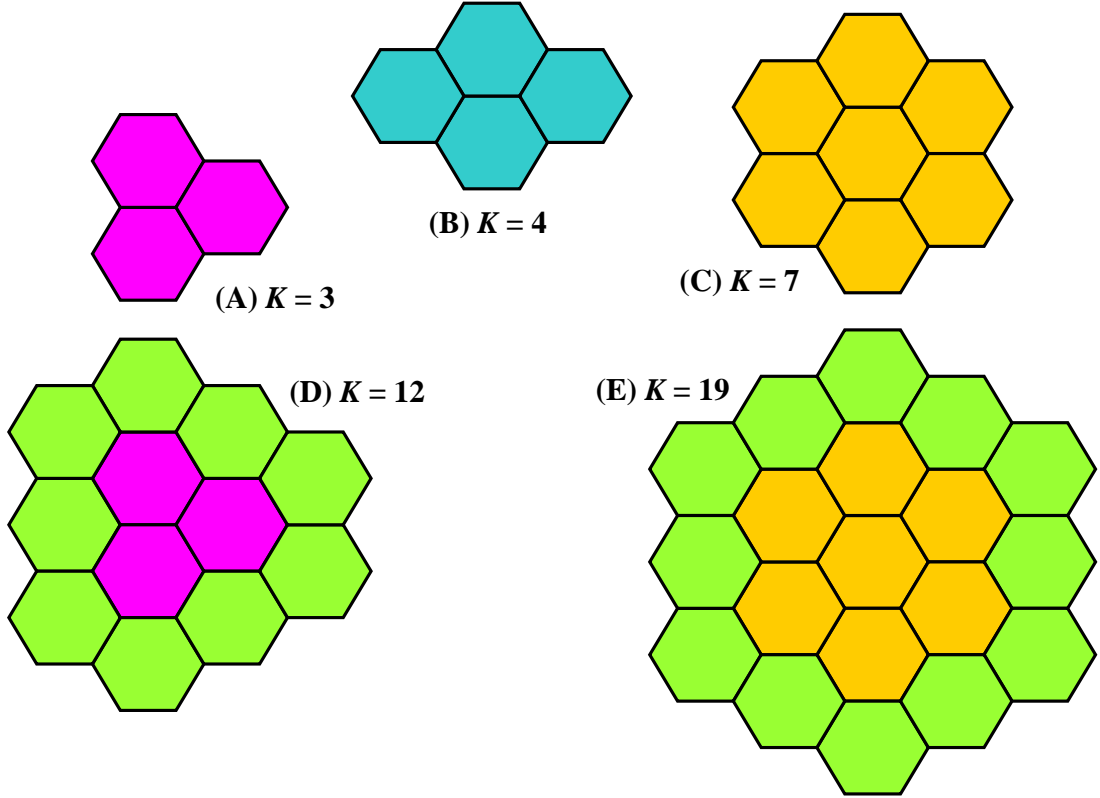}
\caption{Various types of composite detectors consisting of {\it K} closely packed hexagonal encapsulated HPGe detectors. (A) and (B) correspond to different types of miniball detector.}\label{fig:plot1}
\end{figure}

\subsection{Modeling of Cluster detector ({\it K} = 7)}

In a cluster detector, we have two groups of detector modules - six outer detectors and one central detector, as shown in figure 2(A). Let the probability amplitudes for an outer detector be $A', S'_{io}, S'_{oo}, A, S_{io}, S_{oo}$ and the ones for central detector be represented by $A', S'_{ic}, S'_{oc}, A, S_{ic}, S_{oc}$. Note that the absorption probability does not change. Using the general relations $A' + S'_i + S'_o = 1$ and $A + S_i + S_o = 1$, the total events after first interaction are given by
\begin{equation}
T = 7N = N_1 + N_2 
\end{equation}  
where $N_1 = 6N$ and $N_2 = N$. $N_1$ represents the gamma flux that interact with the six outer detectors, and $N_2$ represents the gamma flux interacting with the central detector. Inset of figure 2(A) shows the possible inward scatterings ($S'_i$ and $S_i$) of $\gamma$-ray from different groups of detectors to neighboring detectors. Let us first consider the case of outer detectors.
\begin{equation}
N_1 = N_1S'_{oo} + N_1A' + N_1S'_{io} 
\end{equation}  
From the scattered $\gamma$-rays of an outer detector ($N_1S'_{io}$), $\frac{2}{3}$rd could enter the two adjacent outer detectors while $\frac{1}{3}$rd $\gamma$-rays can enter the central detector, so that
\begin{equation}
S'_{io} = S'_{io}[\frac{2}{3}(A + S_{io} + S_{oo}) + \frac{1}{3}(A + S_{ic} + S_{oc})] 
\end{equation}
\noindent Thus after second interaction, we have
\begin{equation}
N_1 = N_1[\{S'_{oo} + \frac{1}{3}S'_{io}(2S_{oo} + S_{oc})\} + \{A' + S'_{io}A\} + \frac{1}{3}S'_{io}(2S_{io} + S_{ic})] 
\end{equation}
The last term represents scattered $\gamma$-rays. Similar to equation 2.3, we get
\begin{equation}
S_{io} = S_{io}[\frac{2}{3}(A + S_{io} + S_{oo}) + \frac{1}{3}(A + S_{ic} + S_{oc})] 
\end{equation}
The scattered $\gamma$-rays of central detector could enter the outer detectors, so we have
\begin{equation}
S_{ic} = S_{ic}(A + S_{io} + S_{oo}) 
\end{equation}
Thus, after third interaction, we get   
\[ N_1 = N_1[S'_{oo} + \frac{1}{3}S'_{io}\{(2S_{oo} + S_{oc})(1 + \frac{2}{3}S_{io}) + S_{ic}S_{oo}\}]\] 
\begin{equation}
 +~ N_1[A' + S'_{io}A\{1 + \frac{1}{3}(2S_{io} + S_{ic})\}] + \frac{1}{3}N_1S'_{io}[\frac{2}{3}S_{io}(2S_{io} + S_{ic}) + S_{io}S_{ic}]
\end{equation}

\begin{figure}[htp]
\centering
\includegraphics[totalheight=0.49\textheight,viewport=43 260 780 795,clip]{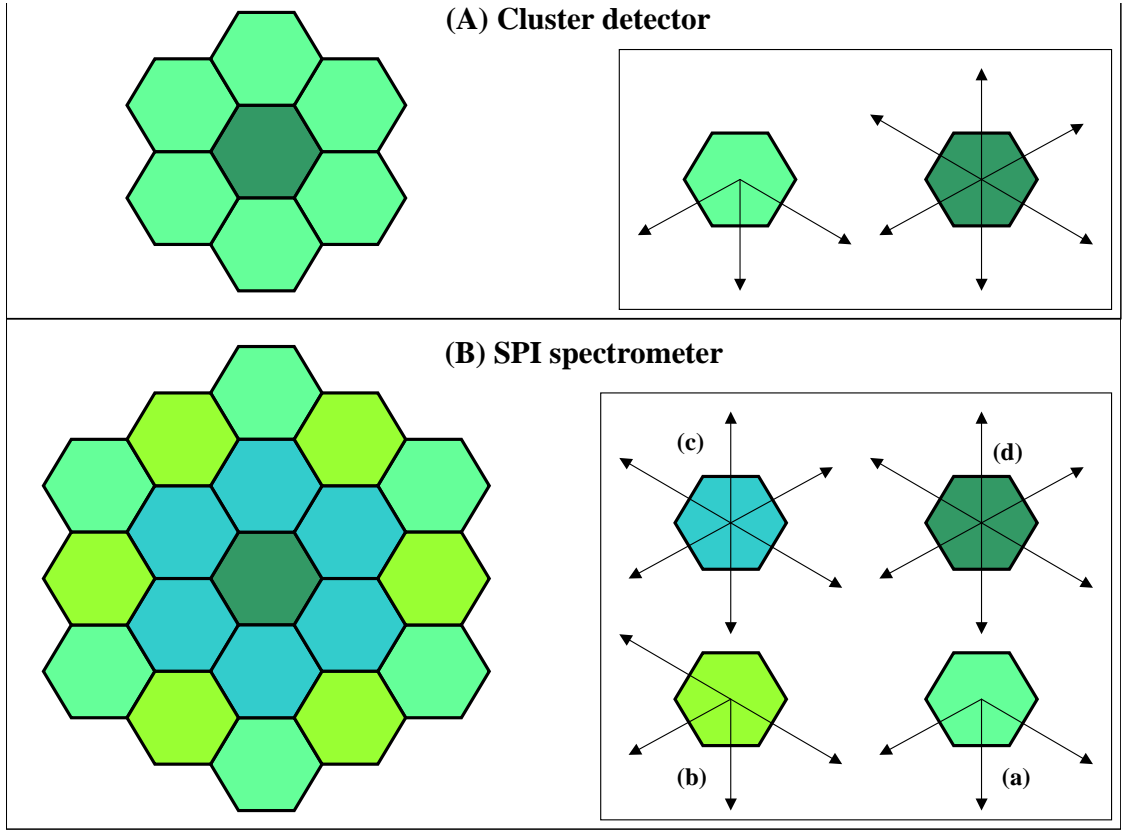}
\caption{The cluster detector and the SPI spectrometer are schematically shown in figures (A) and (B), respectively. For each case, inset shows the possible scatterings ($S'_i$ and $S_i$) of $\gamma$-ray from different groups of detectors to adjacent detectors. Different detector groups are denoted by different colors.}\label{fig:plot1}
\end{figure}

Let us consider the case of central detector.
\begin{equation}
N_2 = N_2S'_{oc} + N_2A' + N_2S'_{ic} 
\end{equation}  
After second interaction, we have   
\begin{equation}
N_2 = N_2S'_{oc} + N_2A' + N_2S'_{ic}(A + S_{io} + S_{oo})  
\end{equation}  
\begin{equation}
~~~~ = N_2[(S'_{oc} + S'_{ic}S_{oo}) + (A' + S'_{ic}A) + S'_{ic}S_{io}] 
\end{equation}  
Using equation 2.5, after third interaction, we get
\begin{equation}
N_2 = N_2[(S'_{oc} + S'_{ic}S_{oo}) + (A' + S'_{ic}A) + S'_{ic}S_{io}\{\frac{2}{3}(A + S_{io} + S_{oo}) + \frac{1}{3}(A + S_{ic} + S_{oc})\}] 
\end{equation}  
Rearranging, we have
\[ N_2 = N_2[S'_{oc} + S'_{ic}S_{oo}(1 + \frac{2}{3}S_{io}) + \frac{1}{3}S'_{ic}S_{io}S_{oc}]\] 
\begin{equation}
~~~~ + N_2[A' + S'_{ic}A(1 + S_{io})] + \frac{1}{3}N_2S'_{ic}S_{io}(2S_{io} + S_{ic})
\end{equation}  
Considering equations 2.7 and 2.12, we observe that the last term consists of four detector interactions and can be neglected. For each case, the first term represents events which escape from detector after partial energy deposition. The partially deposited energy for each case will contribute to background. The second term represents the absorbed events in addback mode. From the symmetry of the detector configuration, it can be observed that if the probability amplitudes for the central detector are $A', S'_i, S'_o, A, S_i, S_o$, then the corresponding amplitudes for each of the outer detectors should be $A', \frac{1}{2}S'_i, (\frac{1}{2}S'_i + S'_o), A, \frac{1}{2}S_i, (\frac{1}{2}S_i + S_o)$. Using these amplitudes and substituting the values of $N_1$ and $N_2$, the expression for total counts becomes
\begin{equation}
T = 7N\beta + 7N(A' + \alpha)
\end{equation} 
where
\begin{equation}
\beta = (S'_o + \frac{3}{7}S'_i) + \frac{1}{7}S'_i(\frac{3}{2}S_i + 4S_{o}) + \frac{1}{7}S'_iS_i(S_i + \frac{5}{2}S_{o})  
\end{equation} 
and 
\[ \alpha = \frac{4}{7}S'_iA + \frac{5}{14}S'_iS_iA\]  
\begin{equation}
 = 0.57~S'_iA + 0.36~S'_iS_iA  .
\end{equation} 
Here the $7N\beta$ counts are due to escaped events and $7N\alpha$ counts can be reconstructed to full energy peak. So, we have,
\begin{equation}
P = 7NA' + 7N\alpha  ,   
\end{equation}                
\begin{equation}
B = 7N\beta .
\end{equation}   
The peak-to-background and peak-to-total ratios are given by
\begin{equation}
P/B = \frac{A' + \alpha}{\beta}  ,
\end{equation}    
\begin{equation}
P/T = A' + \alpha   
\end{equation}
These results are valid even if there are $\gamma$-rays of various energies.
\footnote{If there are $N_1$ $\gamma$-rays of energy $E_1$ and $N_2$ $\gamma$-rays of energy $E_2$ which interact with a detector module, then $P = N_1A'_1 + N_2A'_2, P_{adbk} = N_1(A'_1 + \alpha_1) + N_2(A'_2 + \alpha_2)$ and $T = T = N_1 + N_2$. Thus, the peak-to-total and peak-to-background ratios could be calculated for gamma-rays of various energies.}
It is notable that above result is obtained after neglecting four and higher detector hit events. So, considering equation 2.13, we have in general
\begin{equation}
7N > 7N\beta + 7N(A' + \alpha) 
\end{equation} 
However, if we consider $\gamma$-rays of energy $\approx$ 1.3 MeV, then
\begin{equation}
A' + \alpha + \beta = 1
\end{equation} 
So, the peak-to-background ratio could be written as
\begin{equation}
P/B = \frac{A' + \alpha}{1 - (A' + \alpha)}  ,
\end{equation}    
If we consider the single detector mode, then we realise that corresponding to $N\alpha$ events, the number of counts are $> N\alpha$. Let the corresponding counts be $N(\alpha + \alpha_1)$. Similarly, for $N\beta$ unrecoverable background events, let the corresponding counts be $N(\beta + \beta_1)$. So, the peak-to-background and peak-to-total ratios in single detector mode are given by
\begin{equation}
P_{sd}/B_{sd} = \frac{A'}{(\alpha + \alpha_1) + (\beta + \beta_1)}  ,
\end{equation}    
\begin{equation}
P_{sd}/T_{sd} = \frac{A'}{1 + \alpha_1 + \beta_1}   
\end{equation}
It is important to note that the FEP efficiency in single crystal and addback modes are proportional to $A'$ and $A' + \alpha$, respectively. So, the addback factor ($F$), defined as the ratio of addback efficiency to single detector efficiency is given by 
\begin{equation}
F = \frac{A' + \alpha}{A'}   
\end{equation}

The expressions for peak-to-background and peak-to-total ratios for addback mode depend on $A'$, $\alpha$ and are independent of number of detector modules ({\it K}). These expressions are valid for any composite detector. However, the values will differ depending on values of $A'$ and $\alpha$. $A'$ depends on the shape, size and volume of a detector module. All the composite detectors considered in this text have identical HPGe detector modules. So $A'$ is same for all composite detectors. $\alpha$ is a measure of the capability of the detector system for adding back the scattered events from adjacent detectors to the full energy peak. If there are more detectors surrounding the detector where $\gamma$-ray interacts, then the probability of the $\gamma$-ray to get absorbed increases. So, $\alpha$ depends on the total number of detector modules ({\it K}) constituting the composite detector and will have higher value for a composite detector having more number of detector modules. We will now investigate other composite detectors ({\it K =} 3, 4, 12 and 19) for obtaining detailed expression of $\alpha$.

\subsection{Modeling of SPI spectrometer ({\it K} = 19)}

Let us now consider the SPI spectrometer. From the schematic diagram of the SPI spectrometer in figure 2(B), it can be observed that this composite detector comprises of twelve detector modules surrounding the cluster detector. Compared to cluster detector, this spectrometer consists of approximately three times more number of identical detector modules. From equations 2.19, 2.22 and 2.25, it is observed that knowledge of $\alpha$ is important for calculation of peak-to-total, peak-to-background ratios and addback factor for the addback mode of cluster detector. So for the more sophisticated SPI spectrometer \cite{spi,spi2}, we will use the above formalism to find the expression of $\alpha$. We will consider the absorbed events and events scattered to adjacent detectors, neglecting all the events escaping the detector system. If we consider different inward scatterings of $\gamma$-rays, then the SPI spectrometer can be considered to consist of four groups of detector modules as shown by different colors in inset of figure 2(B). The results for each case are as following:

\begin{enumerate}

\item \underline {Six detector modules of type (a):} Let $N_a$ represent the gamma flux that interact with the six detector modules of type (a), so that, $N_a = 6N$. After first interaction, we have
\begin{equation}
N_a = N_aS'_{oa} + N_aA' + N_aS'_{ia} 
\end{equation}  
From the scattered $\gamma$-rays ($N_aS'_{ia}$), $\frac{2}{3}$rd could enter the two adjacent outer detectors of type b while $\frac{1}{3}$rd $\gamma$-rays can enter a detector of type c, so that
\[ S'_{ia} = S'_{ia}[\frac{2}{3}(A + S_{ib} + S_{ob}) + \frac{1}{3}(A + S_{ic} + S_{oc})]\] 
\begin{equation}
 = S'_{ia}[~A + \frac{1}{3}\{(2S_{ib} + S_{ic}) + (2S_{ob} + S_{oc})\}]
\end{equation}
After second interaction, the absorbed $\gamma$-rays and the ones scattered to adjacent detectors are given by
\begin{equation}
N'_a = N_a[\{A' + S'_{ia}A\} + \frac{1}{3}S'_{ia}(2S_{ib} + S_{ic})] 
\end{equation}
After third interaction, the absorbed $\gamma$-rays are
\begin{equation}
N''_a = N_a[A' + S'_{ia}A + \frac{1}{3}S'_{ia}A(2S_{ib} + S_{ic})] 
\end{equation}

\item \underline {Six detector modules of type (b):} Let $N_b$ represent the gamma flux that interact with the six detector modules of type (b), so that, $N_b = 6N$. After first interaction, we have
\begin{equation}
N_b = N_bS'_{ob} + N_bA' + N_bS'_{ib} 
\end{equation}  
From the scattered $\gamma$-rays, $\frac{1}{2}$ could enter the two adjacent outer detectors of type a while $\frac{1}{2}$ $\gamma$-rays can enter two detectors of type c, so that
\begin{equation}
S'_{ib} = S'_{ib}[\frac{1}{2}(A + S_{ia} + S_{oa}) + \frac{1}{2}(A + S_{ic} + S_{oc})] 
\end{equation}
After second interaction, the absorbed $\gamma$-rays and the ones scattered to adjacent detectors are given by
\begin{equation}
N'_b = N_b[\{A' + S'_{ib}A\} + \frac{1}{2}S'_{ib}(S_{ia} + S_{ic})\}] 
\end{equation}
After third interaction, the absorbed $\gamma$-rays are
\begin{equation}
N''_b = N_b[A' + S'_{ib}A + \frac{1}{2}S'_{ib}A(S_{ia} + S_{ic})] 
\end{equation}

\item \underline {Six detector modules of type (c):} Let $N_c$ represent the gamma flux that interact with the six detector modules of type (c), so that, $N_c = 6N$. After first interaction, we have
\begin{equation}
N_c = N_cS'_{oc} + N_cA' + N_cS'_{ic} 
\end{equation}  
From the scattered $\gamma$-rays, $\frac{1}{3}$rd could enter the detectors of type a and d while $\frac{1}{3}$rd $\gamma$-rays can enter two detectors of type b and c, so that
\begin{equation}
S'_{ic} = S'_{ic}[\frac{1}{6}(A + S_{ia} + S_{oa}) + \frac{1}{6}(A + S_{id} + S_{od}) + \frac{1}{3}(A + S_{ib} + S_{ob}) + \frac{1}{3}(A + S_{ic} + S_{oc})] 
\end{equation}
After second interaction, the absorbed $\gamma$-rays and the ones scattered to adjacent detectors are given by
\begin{equation}
N'_c = N_c[\{A' + S'_{ic}A\} + S'_{ic}\{\frac{1}{3}(S_{ib} + S_{ic}) + \frac{1}{6}(S_{ia} + S_{id})\}] 
\end{equation}
After third interaction, the absorbed $\gamma$-rays are
\begin{equation}
N''_c = N_c[A' + S'_{ic}A + S'_{ic}A\{\frac{1}{3}(S_{ib} + S_{ic}) + \frac{1}{6}(S_{ia} + S_{id})\}] 
\end{equation}

\item \underline {One detector modules of type (d):} Let $N_d$ represent the gamma flux that interact with a detector module of type (d), so that, $N_d = N$. After first interaction, we have
\begin{equation}
N_d = N_dS'_{od} + N_dA' + N_dS'_{id} 
\end{equation}  
All the scattered $\gamma$-rays could enter the six detectors of type c, so that
\begin{equation}
S'_{id} = S'_{id}(A + S_{ic} + S_{oc}) 
\end{equation}
After second interaction, the absorbed $\gamma$-rays and the ones scattered to adjacent detectors are given by
\begin{equation}
N'_d = N_d[\{A' + S'_{id}A\} + S'_{id}S_{ic}] 
\end{equation}
After third interaction, the absorbed $\gamma$-rays are
\begin{equation}
N''_d = N_d[A' + S'_{id}A + S'_{id}AS_{ic}] 
\end{equation}

\end{enumerate}

\noindent From the symmetry of the detector configuration, it can be observed that if the probability amplitudes for the central detector (type d) are $A', S'_i, S'_o, A, S_i, S_o$, then the corresponding amplitudes for detectors of type c should be $A', S'_i, S'_o, A, S_i, S_o$, amplitudes for detectors of type b should be $A', \frac{2}{3}S'_i, (\frac{1}{3}S'_i + S'_o), A, \frac{2}{3}S_i, (\frac{1}{3}S_i + S_o)$ and amplitudes for detectors of type a should be $A', \frac{1}{2}S'_i, (\frac{1}{2}S'_i + S'_o), A, \frac{1}{2}S_i, (\frac{1}{2}S_i + S_o)$. Using these amplitudes and substituting the values of $N_a, N_b, N_c$ and $N_d$, the counts corresponding to absorbed events are given by
\[ N'' = N''_a + N''_b + N''_c + N''_d \]
\begin{equation}
 = 19N[A' + \frac{14}{19}S'_{i}A + \frac{67}{114}S'_{i}AS_i] 
\end{equation}
Thus, for the SPI spectrometer,
\[\alpha = \frac{14}{19}S'_iA + \frac{67}{114}S'_iS_iA\]  
\begin{equation}
 = 0.74~S'_iA + 0.59~S'_iS_iA  ,
\end{equation}

\subsection{Modeling of Miniball detectors ({\it K} = 3, 4)}

\subsubsection{Modeling of three detector configuration}

The three detector configuration for miniball detector is schematically shown in figure 3(A(i)). Let the probability amplitudes for a detector be $a', s'_{i}, s'_{o}, a, s_{i}, s_{o}$. Using the general relations $a' + s'_i + s'_o = 1$ and $a + s_i + s_o = 1$, the total counts after first interaction are given by
\begin{equation}
T = 3N 
\end{equation}  
\begin{equation}
 = 3N(s'_{o} + a' + s'_{i}) 
\end{equation}  
After second interaction, we have
\begin{equation}
T = 3N[s'_o + a' + s'_i(a + s_i + s_o)] 
\end{equation}
\begin{equation}
~~~~~~~~ = 3N(s'_o + s'_is_o) + 3N(a' + s'_ia) + 3Ns'_is_i
\end{equation}
After third interaction, we have   
\begin{equation}
T = 3N(s'_o + s'_is_o) + 3N(a' + s'_ia) + 3Ns'_is_i(a + s_i + s_o)
\end{equation}  
\begin{equation}
  = 3N[s'_o + s'_is_o(1 + s_i)] + 3N[a' + s'_ia(1 + s_i)] + 3Ns'_is_i^2
\end{equation}  
If we neglect the last term corresponding to four detector interactions, and consider the case for cluster detector, we have
$\beta = s'_o + s'_is_o(1 + s_i)$ and $\alpha = as'_i(1 + s_i)$. 

\begin{figure}[htp]
\centering
\includegraphics[totalheight=0.49\textheight,viewport=43 260 780 795,clip]{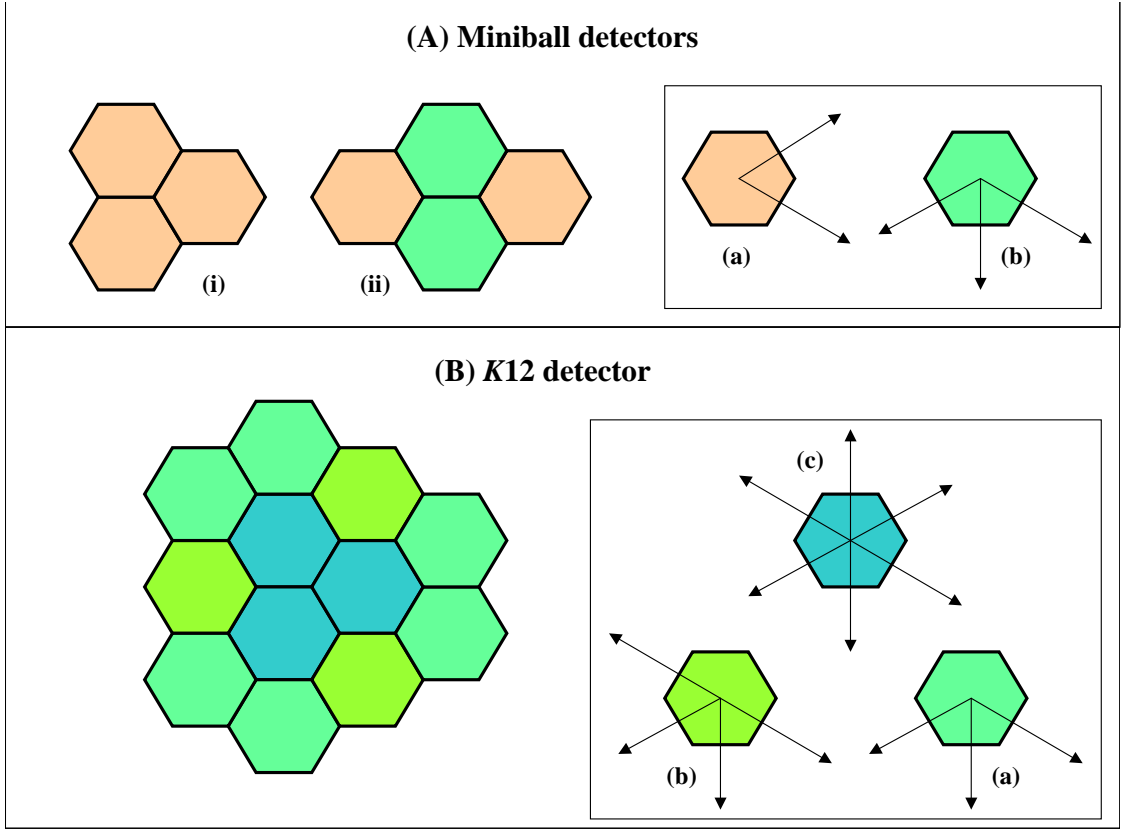}
\caption{The Miniball detector and the {\it K}12 detector are schematically shown in figures (A) and (B), respectively. For each case, inset shows the possible scatterings ($S'_i$ and $S_i$) of $\gamma$-ray from different groups of detectors to adjacent detectors. Different detector groups are denoted by different colors.}\label{fig:plot1}
\end{figure}

\noindent We want a universal treatment for the cluster detector and the miniball detector. From the symmetry of the detector configuration, it can be observed that if the probability amplitudes for the central detector of a cluster detector are $A', S'_i, S'_o, A, S_i, S_o$, then the corresponding amplitudes for a miniball detector module are $A', \frac{1}{3}S'_i, S'_o + \frac{2}{3}S'_i, A, \frac{1}{3}S_i, S_o + \frac{2}{3}S_i$. So, we have
\begin{equation}
\alpha = \frac{1}{3}S'_iA + \frac{1}{9}S_iS'_iA
\end{equation}

\subsubsection{Modeling of four detector configuration}

The three detector configuration for miniball detector is schematically shown in figure 3(A(ii)). This detector can be considered to consist of two groups of detector modules as shown by different colors of figure 3(A(ii)). The results for each case are as following:

\begin{enumerate}

\item \underline {Two detector modules of type (a):} Let $N_a$ represent the gamma flux that interact with the two detector modules of type (a), so that, $N_a = 2N$. After first interaction, we have
\begin{equation}
N_a = N_aS'_{oa} + N_aA' + N_aS'_{ia} 
\end{equation}  
All the scattered $\gamma$-rays ($N_aS'_{ia}$) could enter adjacent detectors of type b, so that
\begin{equation}
S'_{ia} = S'_{ia}(A + S_{ib} + S_{ob}) 
\end{equation}
After second interaction, the absorbed $\gamma$-rays and the ones scattered to adjacent detectors are given by
\begin{equation}
N'_a = N_a[\{A' + S'_{ia}A\} + S'_{ia}S_{ib}] 
\end{equation}
After third interaction, the absorbed $\gamma$-rays are
\begin{equation}
N''_a = N_a[A' + S'_{ia}A + S'_{ia}S_{ib}A] 
\end{equation}

\item \underline {Two detector modules of type (b):} Let $N_b$ represent the gamma flux that interact with the two detector modules of type (b), so that, $N_b = 2N$. After first interaction, we have
\begin{equation}
N_b = N_bS'_{ob} + N_bA' + N_bS'_{ib} 
\end{equation}  
From the scattered $\gamma$-rays, $\frac{2}{3}$rd could enter adjacent detectors of type a and $\frac{1}{3}$rd $\gamma$-rays could enter adjacent detector of type b, so that
\[ S'_{ib} = S'_{ib}\frac{1}{3}[2(A + S_{ia} + S_{oa}) + (A + S_{ib} + S_{ob})]\] 
\begin{equation}
 = S'_{ib}[~A + \frac{1}{3}\{(2S_{ia} + S_{ib}) + (2S_{oa} + S_{ob})\}]
\end{equation}
After second interaction, the absorbed $\gamma$-rays and the ones scattered to adjacent detectors are given by
\begin{equation}
N'_b = N_b[\{A' + S'_{ib}A\} + \frac{1}{3}S'_{ib}(2S_{ia} + S_{ib})] 
\end{equation}
After third interaction, the absorbed $\gamma$-rays are
\begin{equation}
N''_b = N_b[A' + S'_{ib}A + \frac{1}{3}S'_{ib}A(2S_{ia} + S_{ib})] 
\end{equation}

\end{enumerate}

\noindent From the symmetry of the detector configuration, it can be observed that if the probability amplitudes for the central detector of a cluster detector are $A', S'_i, S'_o, A, S_i, S_o$, then the corresponding amplitudes for detectors of type b should be $A', \frac{1}{2}S'_i, (\frac{1}{2}S'_i + S'_o), A, \frac{1}{2}S_i, (\frac{1}{2}S_i + S_o)$ and amplitudes for detectors of type a should be $A', \frac{1}{3}S'_i, (\frac{1}{3}S'_i + S'_o), A, \frac{1}{3}S_i, (\frac{1}{3}S_i + S_o)$. Using these amplitudes and substituting the values of $N_a, N_b$ and $N_c$, the counts corresponding to absorbed events are given by
\[ N'' = N''_a + N''_b\]
\begin{equation}
 = 4N[A' + \frac{5}{12}S'_{i}A + \frac{13}{72}S'_{i}S_iA] 
\end{equation}
Thus, for the miniball four detector configuration,
\begin{equation}
\alpha = \frac{5}{12}S'_{i}A + \frac{13}{72}S'_{i}S_iA  
\end{equation}

\subsection{Modeling of {\it K}12 detector ({\it K} = 12)}

This detector can be considered to consist of three groups of detector modules as shown by different colors of figure 3(B). The results for each case are as following:

\begin{enumerate}

\item \underline {Six detector modules of type (a):} Let $N_a$ represent the gamma flux that interact with the six detector modules of type (a), so that, $N_a = 6N$. After first interaction, we have
\begin{equation}
N_a = N_aS'_{oa} + N_aA' + N_aS'_{ia} 
\end{equation}  
From the scattered $\gamma$-rays ($N_aS'_{ia}$), $\frac{1}{3}$rd could enter adjacent detectors of type a, b and c, so that
\[ S'_{ia} = S'_{ia}\frac{1}{3}[(A + S_{ia} + S_{oa}) + (A + S_{ib} + S_{ob}) + (A + S_{ic} + S_{oc})]\] 
\begin{equation}
 = S'_{ia}[~A + \frac{1}{3}\{(S_{ia} + S_{ib} + S_{ic}) + (S_{oa} + S_{ob} + S_{oc})\}]
\end{equation}
After second interaction, the absorbed $\gamma$-rays and the ones scattered to adjacent detectors are given by
\begin{equation}
N'_a = N_a[\{A' + S'_{ia}A\} + \frac{1}{3}S'_{ia}(S_{ia} + S_{ib} + S_{ic})] 
\end{equation}
After third interaction, the absorbed $\gamma$-rays are
\begin{equation}
N''_a = N_a[A' + S'_{ia}A + \frac{1}{3}S'_{ia}A(S_{ia} + S_{ib} + S_{ic})] 
\end{equation}

\item \underline {Three detector modules of type (b):} Let $N_b$ represent the gamma flux that interact with the three detector modules of type (b), so that, $N_b = 3N$. After first interaction, we have
\begin{equation}
N_b = N_bS'_{ob} + N_bA' + N_bS'_{ib} 
\end{equation}  
From the scattered $\gamma$-rays, $\frac{1}{2}$ could enter the two adjacent outer detectors of type a and c, so that
\begin{equation}
S'_{ib} = S'_{ib}[\frac{1}{2}(A + S_{ia} + S_{oa}) + \frac{1}{2}(A + S_{ic} + S_{oc})] 
\end{equation}
After second interaction, the absorbed $\gamma$-rays and the ones scattered to adjacent detectors are given by
\begin{equation}
N'_b = N_b[\{A' + S'_{ib}A\} + \frac{1}{2}S'_{ib}(S_{ia} + S_{ic})\}] 
\end{equation}
After third interaction, the absorbed $\gamma$-rays are
\begin{equation}
N''_b = N_b[A' + S'_{ib}A + \frac{1}{2}S'_{ib}A(S_{ia} + S_{ic})] 
\end{equation}

\item \underline {Three detector modules of type (c):} Let $N_c$ represent the gamma flux that interact with the three detector modules of type (c), so that, $N_c = 3N$. After first interaction, we have
\begin{equation}
N_c = N_cS'_{oc} + N_cA' + N_cS'_{ic} 
\end{equation}  
From the scattered $\gamma$-rays, $\frac{1}{3}$rd could enter adjacent detectors of type a, b and c, so that
\[ S'_{ic} = S'_{ic}\frac{1}{3}[(A + S_{ia} + S_{oa}) + (A + S_{ib} + S_{ob}) + (A + S_{ic} + S_{oc})]\] 
\begin{equation}
 = S'_{ic}[~A + \frac{1}{3}\{(S_{ia} + S_{ib} + S_{ic}) + (S_{oa} + S_{ob} + S_{oc})\}]
\end{equation}
After second interaction, the number of absorbed $\gamma$-rays and $\gamma$-rays scattered to adjacent detectors are given by
\begin{equation}
N'_c = N_c[\{A' + S'_{ic}A\} + \frac{1}{3}S'_{ic}(S_{ia} + S_{ib} + S_{ic})] 
\end{equation}
After third interaction, the number of absorbed $\gamma$-rays are
\begin{equation}
N''_c = N_c[A' + S'_{ic}A + \frac{1}{3}S'_{ic}A(S_{ia} + S_{ib} + S_{ic})] 
\end{equation}

\end{enumerate}

\noindent From the symmetry of the detector configuration, it can be observed that if the probability amplitudes for the central detector (type c) are $A', S'_i, S'_o, A, S_i, S_o$, then the corresponding amplitudes for detectors of type b should be $A', \frac{2}{3}S'_i, (\frac{1}{3}S'_i + S'_o), A, \frac{2}{3}S_i, (\frac{1}{3}S_i + S_o)$ and amplitudes for detectors of type a should be $A', \frac{1}{2}S'_i, (\frac{1}{2}S'_i + S'_o), A, \frac{1}{2}S_i, (\frac{1}{2}S_i + S_o)$. Using these amplitudes and substituting the values of $N_a, N_b$ and $N_c$, the counts corresponding to absorbed events are given by
\[ N'' = N''_a + N''_b + N''_c \]
\begin{equation}
 = 12N[A' + \frac{2}{3}S'_{i}A + \frac{11}{24}S'_{i}S_iA] 
\end{equation}
Thus, for the {\it K}12 detector,
\begin{equation}
\alpha = \frac{2}{3}S'_{i}A + \frac{11}{24}S'_{i}S_iA  
\end{equation}

\section{Results and discussion}

\subsection{Comparison of observations}

Let us consider the addback mode of a composite detector having {\it K} detector modules. For a $\gamma$-ray of energy $E_{\gamma}$, if the number of single, double and triple fold events contributing to FEP be $Kh_{1}, Kh_{2}$ and $Kh_{3}$, then
\begin{equation}
h_{1} = A' 
\end{equation} 
\begin{equation}
h_{2} = \rho_1 S'_iA
\end{equation} 
\begin{equation}
h_{3} = \rho_2 S'_iAS_i
\end{equation}
where the coefficients $\rho_1$ and $\rho_2$ depend on the configuration of detectors constituting the composite detector. So, $\alpha$ could be written as
\begin{equation}
\alpha = h_2 + h_3  
\end{equation}
Now let us define quantities in order to extract the fold distribution. If the ratio of single fold events to total full energy peak events be denoted by $h'_{1}$, the ratio of double events to total full energy peak events be denoted by $h'_2$ and the ratio of triple events to total full energy peak events be denoted by $h'_3$, then
\begin{equation}
h'_{j} = \frac{h_j}{h_1 + h_2 + h_3}
\end{equation} 
where j = 1, 2 and 3, corresponding to single, double and triple fold events. The values of $h'_j (j = 1, 2, 3)$ for a particular gamma-ray energy gives us information about the fold distribution at that energy. Considering equations 2.19 and 2.22, we infer that peak-to-total and peak-to-background ratios could be extracted if we know the values of $h'_j (j = 1, 2, 3)$. The fold distribution helps in understanding the behavior of peak-to-total and peak-to-background ratios.

Let us consider the expressions of $\alpha$ for the five types of composite detectors discussed above. Considering equations 2.15, 2.43, 2.50, 2.60 and 2.74, the values of $\rho_1$ and $\rho_2$ could be extracted. These values of $\rho_1$ and $\rho_2$ have been plotted as a function of number of detector modules ({\it K}) in figure 4. It is observed that with increasing value of {\it K}, the values of $\rho_1$ and $\rho_2$ gradually increase. This is due to more contribution of multiple hit events for composite detector having higher number of detector modules. 

\noindent In general, we have $\alpha_{Miniball({\it K =}3)} < \alpha_{Miniball({\it K =}4)} < \alpha_{cluster({\it K =}7)} < \alpha_{SPI({\it K =}19)}$. As a result, from equation 2.21 we have $\beta_{Miniball({\it K =}3)} > \beta_{Miniball({\it K =}4)} > \beta_{cluster({\it K =}7)} > \beta_{SPI({\it K =}19)}$. Considering all these detectors (equations 2.15, 2.43, 2.50, 2.60, 2.74 and 2.19), we have
\begin{equation}
\frac{P}{T}(cluster) - \frac{P}{T}(miniball(3)) = 0.24~S'_iA + 0.25~S'_iS_iA  ,
\end{equation} 
\begin{equation}
\frac{P}{T}(cluster) - \frac{P}{T}(miniball(4)) = 0.15~S'_iA + 0.18~S'_iS_iA  ,
\end{equation} 
\begin{equation}
\frac{P}{T}(SPI) - \frac{P}{T}(cluster) = 0.17~S'_iA + 0.23~S'_iS_iA  ,
\end{equation} 
\begin{equation}
\frac{P}{T}({\it K}12) - \frac{P}{T}(cluster) = 0.10~S'_iA + 0.10~S'_iS_iA  ,
\end{equation} 
Thus, increased peak counts and lower background (including escape peak counts) in addback mode causes a better performance of the cluster detector compared to the miniball detectors and the SPI spectrometer compared to the cluster detector.

\begin{figure}[htp]
\centering
\includegraphics[totalheight=0.49\textheight,viewport=43 260 780 795,clip]{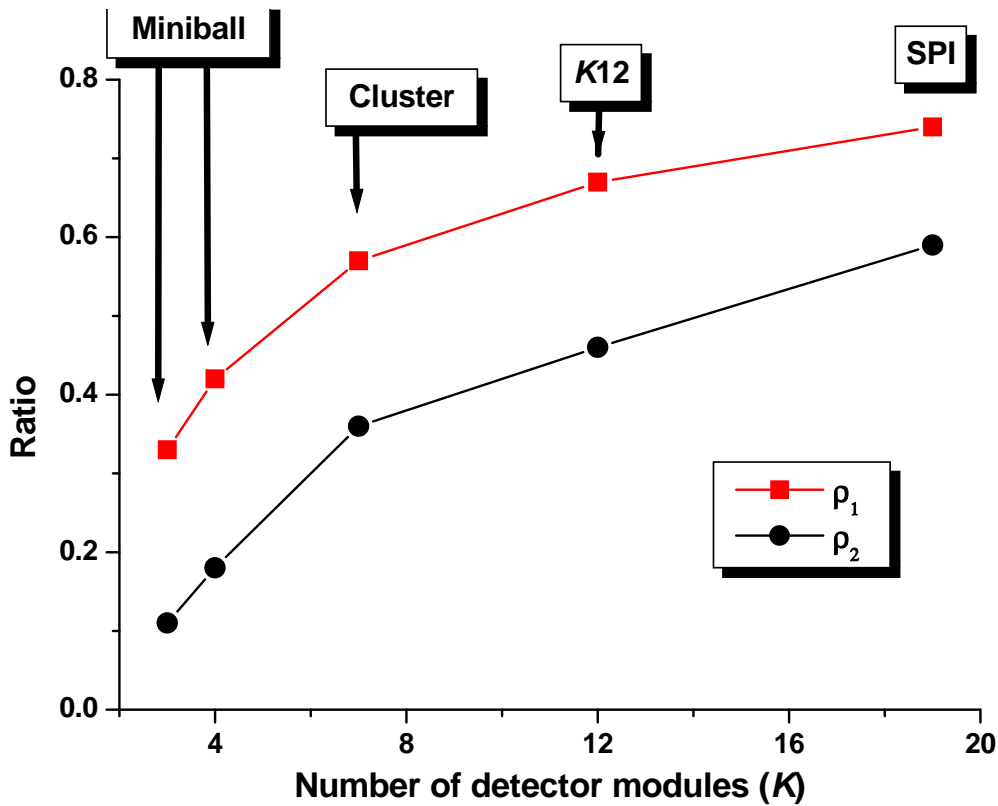}
\caption{$\rho_1$ and $\rho_2$ have been plotted as a function of number of detector modules ({\it K}).}\label{fig:plot1}
\end{figure}

\subsection{Predictions using experimental data of cluster detector}

The present work attempts to undertake from first principles the analysis of a complex detection problem. A detailed formalism has been presented for understanding the various modes of operation of a composite detector. Regarding practical usefulness of this formalism, we could calculate the values of all probability coefficients from considerations of cross-sections for the various gamma-ray interaction processes with matter. This is a way of comparing the outcomes of this formalism with experimental data. Another interesting possibility is - we consider the experimental data of a composite detector as input, thereafter using the present formalism, we can predict and compare the response of other composite detectors. This can help us to get insight about their operation and could provide a guidance for experimental studies with other composite detectors. We will now investigate this possibility using known information about the cluster detector \cite{wil}.

For a $\gamma$-ray of energy $E_{\gamma}$, if the number of double fold events to single fold events (both contributing to FEP) be denoted by $h_{21}$ and the number of triple fold events to double fold events (both contributing to FEP) be denoted by $h_{32}$, then from equations 2.13 and 2.15, we have 
\begin{equation}
h_{21} = \frac{4}{7}\frac{S'_iA}{A'}
\end{equation} 
and
\begin{equation}
h_{32} = \frac{5}{8}S_i
\end{equation} 
The experimental fold distribution of the cluster detector for 1.3 MeV $\gamma$-ray \cite{wil} shows that $h'_1 \approx 0.66, h'_2 \approx 0.30$ and $h'_3 \approx 0.04$. Considering these values and equation 3.10, we get $S_i$ = 0.21. Similarly from equation 3.9, we have $S'_iA = 0.80A'$. The experimental value of peak-to-total ratio in addback mode at 1332 keV is found to be $\approx$ 0.39 \cite{ebe}. Using equations 2.15, 2.19 and above information, we get $A' = 0.26$. Thus, we have
\begin{equation}
h_{1} = 0.26 
\end{equation} 
\begin{equation}
h_{2} = 0.21\times \rho_1
\end{equation} 
\begin{equation}
h_{3} = 0.04\times \rho_2
\end{equation}
For the different types of composite detectors considered in this text, we can now use the value of $\rho_1$ and $\rho_2$ along with above equations for extracting the fold distribution, relative FEP efficiency, peak-to-total ratio and peak-to-background ratio. The results of fold distribution for various composite detectors ({\it K = }3, 4, 7, 12 and 19) is shown for 1332 keV gamma-ray in figure 5. We observe that the single fold events are dominant. The number of multiple fold events - in particular two fold events - increases slowly with increasing value of {\it K}. The amount of three fold events is small for {\it K} = 3 to 19, varying from 1.5\% to 5.8\%. So, the double fold coincidences play an important role for the add-back mode.

\begin{figure}[htp]
\centering
\includegraphics[totalheight=0.49\textheight,viewport=43 260 780 795,clip]{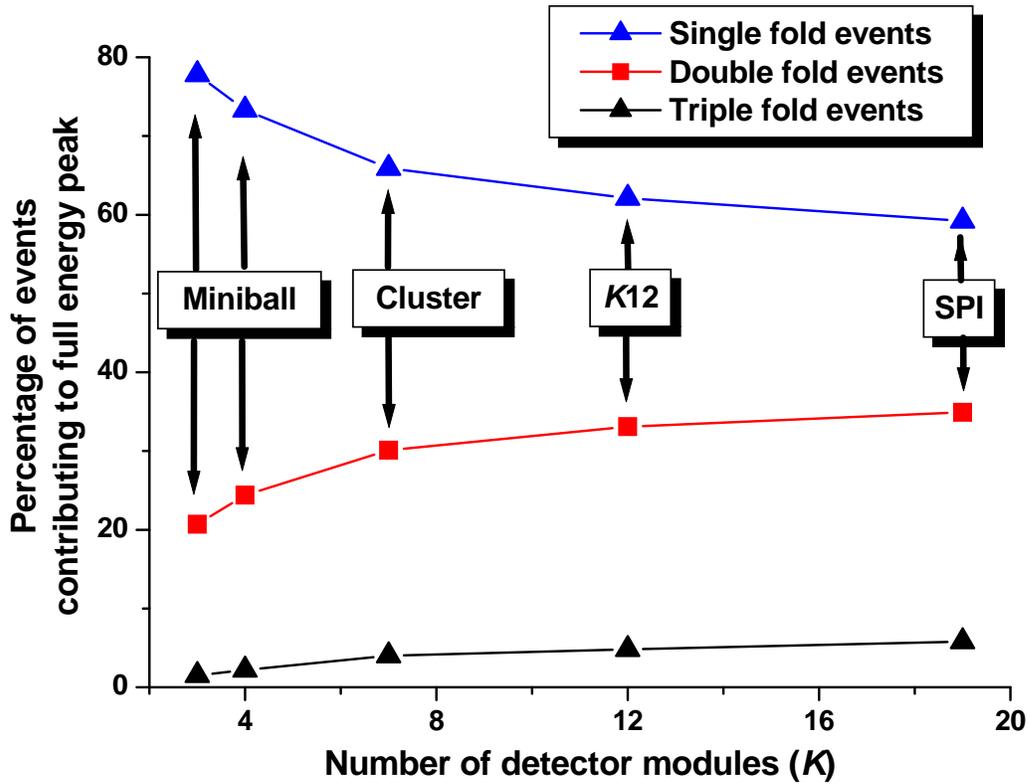}
\caption{Fold distribution have been shown as a function of number of detector modules ({\it K}) for 1332 keV $\gamma$-ray.}\label{fig:plot1}
\end{figure}

The relative FEP efficiency in single detector and addback modes are proportional to ${\it K}h_1$ and ${\it K}(h_1 + h_2 + h_3)$, respectively. A comparison of relative FEP efficiencies (at 1332 keV) for different composite detectors have been shown in figure 6(A). Results have been shown for both single detector and addback modes of operation. Compared to single detector mode, the increase in values for addback mode is due to the increasing contribution of multiple fold events.

\begin{figure}[htp]
\centering
\includegraphics[totalheight=0.75\textheight,viewport=40 60 550 800,clip]{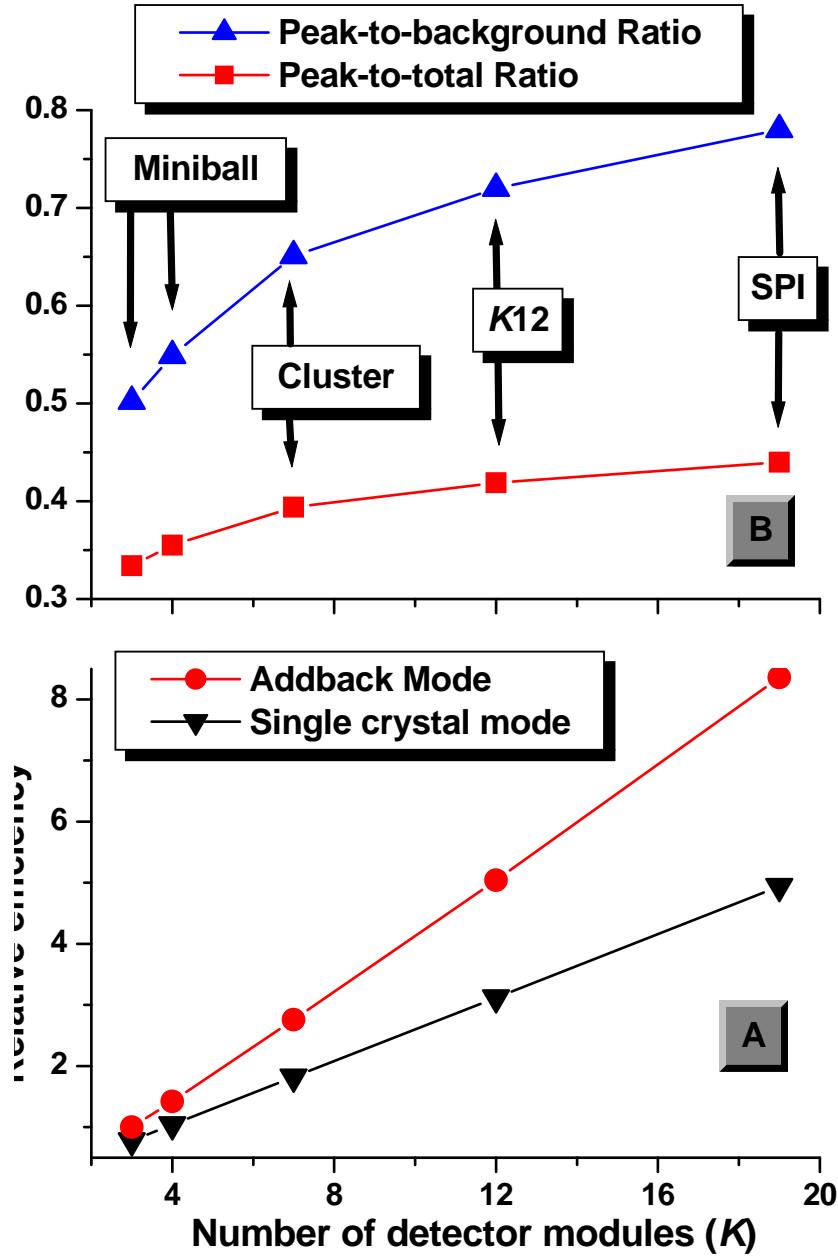}
\caption{(A) shows relative FEP efficiency (at 1332 keV) for different modes of operation as a function of number of detector modules ({\it K}). Peak-to-total and peak-to-background ratios in addback mode have been plotted for various composite detectors in (B).}\label{fig:hitp}
\end{figure}

Peak-to-total and peak-to-background ratios in addback mode have been calculated as a function of number of detector modules ({\it K}) and the results have been shown for 1332 keV gamma-ray in figure 6(B). Both ratios show an increasing trend with higher value of {\it K}. This is due to increase of multiple fold events ($h_2$ and $h_3$), as observed from figure 5. Compared to peak-to-background ratio, the peak-to-total ratio increases at a slower rate. The value of peak-to-total and peak-to-background ratios of the SPI spectrometer at 1332 keV are found to be $\approx$ 0.44 and $\approx$ 0.79, respectively.

\subsection{Comparison of predictions and experimental results for SPI spectrometer}

Attie {\it  et al.} have performed FEP measurements of the SPI spectrometer \cite{spi2}. From the measurements (table 1 and 2 of \cite{spi2}), it is observed that for 1332 keV $\gamma$-ray, the single event (SE) efficiency (or single detector efficiency) and multiple event (ME) efficiency are 16.1$\%$ and 10.0$\%$, respectively. Note that the FEP efficiency in addback mode is sum of SE and ME efficiencies. So, the addback factor (defined by equation 2.25) at 1332 keV is $\frac{16.1 + 10.0}{16.1} = 1.62$. From the predictions for SPI spectrometer of previous section (figure 6(A)), we observe that the value of addback factor at 1332 keV is $\frac{8.36}{4.94} = 1.69$, which is quite close to the experimental value of 1.62. From figure 4 of \cite{spi2}, we observe that at 1332 keV, two fold and three fold efficiencies are $\approx$ 8.5 and 1.5 $\%$, respectively. Thus, for 1332 keV $\gamma$-ray, the experimental ratio of single, double and triple fold events is $16.1 : 8.5 : 1.5 = 10.7 : 5.7 : 1.0$. From the present formalism, the value of this ratio for 1332 keV $\gamma$-ray (figure 5) is $59.2 : 34.9 : 5.8 = 10.2 : 6.0 : 1.0$. Again we observe remarkable agreement between the experimental data of Attie {\it  et al.} \cite{spi2} and the results from the present formalism. From these two agreements, we conclude that the present formalism successfully describes the response of a general composite detector including sophisticated spectrometer like SPI. 

The predictions of the present formalism are based on fact that the source is placed axially at a distance of $\approx$ 25 from the detector \cite{ebe,wil}. However, in reality the gamma-rays from outer space come from sources which are practically at infinity compared to 25 cm. So, we have parallel gamma-rays falling on the SPI spectrometer. For large source-to-detector distances, a gamma-ray incident on the detector will encounter approximately uniform thickness of the detector active volume. However, for moderate distances, the variation of thickness seen by the gamma-ray is considerable. The effect of this edge penetration could be seen in the detector efficiency curves shown by Bell \cite{bell}. Due to detector edge effects, there is an increase in FEP efficiency for extremely large source-to-detector distances compared to the FEP efficiency for source-to-detector distance $\approx$ 25 cm. As a result, for the SPI spectrometer, the values of relative FEP efficiency and peak-to-total, peak-to-background ratios shown in figures 6(A) and 6(B) are lower compared to the actual values. Even for decreased value of efficiencies compared to actual ones, we have seen from the previous paragraph that the ratio of efficiencies from the present formalism agree nicely with the experimental data of Attie {\it  et al.} \cite{spi2}. So, the present formalism helps in getting estimates of the fold distribution and the peak-to-total, peak-to-background ratios of complex spectrometer like SPI, for the first time through an indirect approach at 1332 keV.

\section{Summary and conclusion}

The present work attempts to undertake from first principles the analysis of a complex detection problem and presents a way of understanding various types of composite detectors using experimental data. Based on absorption and scattering of $\gamma$-rays, a probability model has been presented for understanding the addback mode of operation for the miniball detector, the cluster detector, the SPI spectrometer and a composite detector comprising of fourteen encapsulated HPGe detectors. In the present formalism, the operation of these sophisticated detectors could be described in terms of six probability amplitudes only. Considering up to triple detector hit events, we have obtained expressions for fold distribution, peak-to-background and peak-to-total ratios for addback mode of operation of the composite detectors. Results indicate improved performance of the SPI spectrometer compared to other composite detector in addback mode. Using experimental data on relative efficiency and fold distribution of cluster detector as input, the fold distribution and the peak-to-total, peak-to-background ratios have been calculated for the SPI spectrometer and other composite detectors at 1332 keV. 
For the SPI spectrometer, remarkable agreement between the experimental data and the results from the present formalism, has been observed. This demonstrates the power of the formalism to predict the response of composite detectors similar to the cluster detector. These results could be helpful for providing a guidance in experimental studies with sophisticated spectrometer like SPI.


\end{document}